\documentclass[runningheads]{llncs}
\usepackage{cite}   
\usepackage{booktabs}   
\usepackage{array}      
\usepackage{subfig} 
\usepackage{array, booktabs, makecell}   
\usepackage{tabularx}                    
\usepackage{multirow}   
\usepackage{makecell}   
\usepackage{float}
\usepackage{subcaption}
\usepackage{ragged2e}  
\usepackage{silence}
\usepackage{enumitem}

\usepackage[T1]{fontenc}
\usepackage{graphicx}
\makeatletter
\AtBeginDocument{%
  \renewcommand*\caption@Warning[1]{}%
}
\makeatother

\begin{document}
\title{Co-Designing Digital Humans for Online Learning: A Framework for Human-AI Pedagogical Integration}
\titlerunning{Co-Designing Digital Humans for Online Learning}
%
%
\author{Xiaokang Lei\inst{1} \orcidID{0009-0004-4158-293X} 
\and
Ching Christie Pang\inst{2} \orcidID{0000-0003-4704-2403} 
\and
Yuyang Jiang\inst{2} \orcidID{0009-0005-3571-6935}  
\and
Xin Tong\inst{1} \orcidID{0000-0002-8037-6301}  
\and
Pan Hui\inst{1,2} \orcidID{0000-0001-6026-1083}}
\authorrunning{X. Lei et al.}
%
\institute{Hong Kong University of Science and Technology (Guangzhou), Guangzhou, China 
Hong Kong University of Science and Technology, Hong Kong SAR, China
}
\maketitle              

\begin{figure}[H] 
  \centering
  \vspace{-2.8em} 
  \includegraphics[width=\textwidth, trim=0 5 0 5, clip]{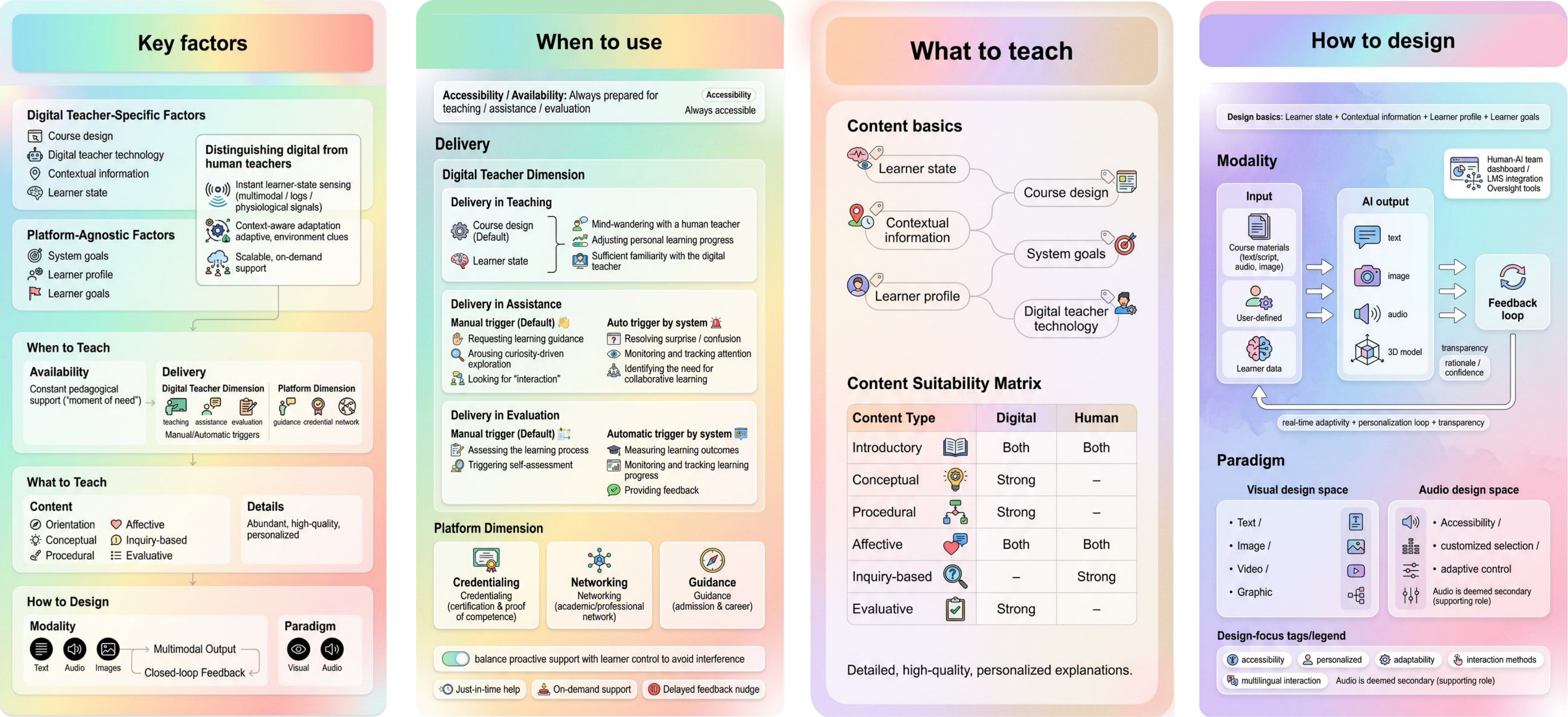} 
  \vspace{-1em} 
  \caption{An overview of our proposed framework. From left to right: an example of key factors and design space in adopting digital teachers in online learning; including when to use; what to teach; and how to design.}
  \label{fig:overview}
  \vspace{-2em} 
\end{figure}

\begin{abstract}
Artificial intelligence (AI) and large language models (LLMs) are reshaping education, with virtual avatars emerging as digital teachers capable of enhancing engagement, sustaining attention, and addressing instructor shortages. Aligned with the Sustainable Development Goals (SDGs) for equitable quality education, these technologies hold promise yet lack clear guidelines for effective design and implementation in online learning. To fill this gap, we introduce a framework specifying when, what, and how digital teachers should be integrated. Our study combines (1) a design space analysis of 87 works across AI, educational technology, design, and HCI, (2) a survey of 132 learners' practices and preferences, and (3) three co-design workshops with 18 experts from pedagogy, design, and AI. It provides actionable guidance for educators, designers, and HCI researchers, advancing opportunities to build more engaging, equitable, and effective online learning environments powered by digital teachers.

\keywords{\textbf{Digital Teachers \and Higher Education \and Hybrid Learning \and Large-Scale Deployment \and University Students \and Human-Computer Interaction.}}

\end{abstract}

\section{Introduction}
Artificial intelligence (AI) and large language models (LLMs) are reshaping online learning. Among the visible developments are digital humans, the AI‑powered avatars or "digital teachers" that deliver instruction through conversation, embodiment, and adaptive behavior \cite{Schroeder2021LearningVirtualHumans, Pang2025}. These agents can leverage embodied intelligence to enhance engagement, sustain attention, and address teacher shortages \cite{Fu2025, Tan2025}. A major advantage of digital human technologies is their high scalability \cite{oliveira2024unveiling, weeraskara2024digital}, allowing easy deployment across platforms and enabling educational access at scale. 

AI‑driven tutoring systems have been shown to improve learner satisfaction and performance over traditional methods \cite{yu2024enhancing, basri2024effectiveness, jothikumar2025ai}. With this strong advantage, HCI researchers and educators are now studying this technology as it shapes a new, accessible form of pedagogy \cite{Tan2025, Fu2025, Pang2025}. Furthermore, digital teachers can support personalization, adapting content to meet individual learner needs and preferences \cite{Tan2025}. Despite their promise, however, clear guidelines for effective design and pedagogical integration with the emerging technologies remain scarce \cite{Tan2025, Pang2025}. 

"Learning anytime and anywhere" has gained traction in recent years, particularly with the rise of Massive Open Online Courses (MOOCs) and online learning \cite{du2022learning, fletcher2007learning}. It has democratized global access to education \cite{pappano2012year}. However, incorporating AI and digital teachers introduces new challenges that require careful consideration \cite{alenezi2023need, pedro2019artificial}. While the HCI community has explored AI in education, a practical 
framework 
specifying when, what, and how digital humans should intervene has not been established.

To address these gaps, we developed a co‑design framework collaboratively created by technicians, educators, instructional designers, and HCI researchers. The framework guides decisions on the timing, content, and design of digital teacher interventions in online courses. It conceptualizes digital humans as educational partners that complement human instructors rather than replace them.

This framework advances the HCI vision of creating context‑aware, personalized, and pedagogically meaningful AI systems. By formalizing how teachers, designers, and researchers can collaboratively shape digital humans\cite{yu2024enhancing,basri2024effectiveness,jothikumar2025ai}, it promotes inclusive and adaptive online learning environments aligned with the UN Sustainable Development Goal of "Quality Education for All" \cite{ref_sdg4}. 

Our co-design framework aimed at guiding the integration of digital teachers into online learning environments. Our study involved a \textbf{three-phase method}: (i) a design space analysis of 87 works and studies across AI, education, design, and HCI to identify design space and key factors; (ii) a survey with 132 learners on their practices and preferences; (iii) three co-design workshops with 18 interdisciplinary experts to develop an actionable framework. Through our research, we aim to answer two critical questions: 
\textbf{RQ1:} What are the advantages and disadvantages of having a digital teacher involved in the online learning process (e.g., MOOCs)? What are the attitudes of learners and teachers towards the integration of AI digital humans? 
\textbf{RQ2:} What are the design space, key factors and framework 
for effective online learning and AI digital teachers integration? 

Our findings reveal several interesting insights. \textbf{First}, learners expressed a strong preference for personalized interactions with digital teachers, indicating that they value tailored feedback and support. This suggests that the design of digital teachers should prioritize adaptability and responsiveness to individual learner needs. \textbf{Second}, our workshops with experts highlighted the importance of learner state in guiding AI systems, with participants emphasizing the need for the system to automatically address and understand learner's actions and states in order to provide the corresponding education services. This aligns with existing literature on the importance of customization and personalization in AI learning systems \cite{kaswan2024ai, murtaza2022ai, ayeni2024ai}.

Through our findings, we aim to contribute to the discourse on AI in education by presenting a structured \textbf{design space} that encompasses \textbf{key factors}, including \textbf{what to teach}, \textbf{when to use} digital teachers, and \textbf{how to design} these interactions. 
The contributions of this research are:

    (1) We developed the first framework specifically tailored for the integration of digital teachers into online education.
    
    (2) We summarized literature from multiple domains and identified key dimensions addressing the \textbf{when, what, and how} to integrate digital teachers within the intersection of education, AI digital teachers, and online learning platforms.
    
    (3) Our study involving 132 learners revealed current user dissatisfaction with online learning (e.g., MOOCs) and digital teachers, providing future directions for HCI researchers to address these challenges.
    
    (4) The outcomes of co-design workshops with 18 experts demonstrated that the proposed framework provides valuable insights and support for educators, facilitating the development of effective strategies for utilizing digital human technologies in teaching.

\section{Related Work} \label{sec:rw}

This section reviews the background of educational digital teachers and existing online learning frameworks to demonstrate the need for a new, integrated framework for digital teacher-driven education.

\subsection{Digital Human in Education}

Digital humans (or virtual agents) are user-centric entities simulating human appearance and conversation to facilitate learning \cite{lee2024, Swartout2006TowardVirtualHumans, Schroeder2021LearningVirtualHumans}. Evolution from problem-solving tutoring systems \cite{Lane2022} to LLM-powered entities has enabled real-world deployments \cite{Pang2025} and humanized interactions \cite{Zhang2025}.

Research focuses on design features and technical integration. Avatar appearance significantly impacts satisfaction and learning experiences \cite{Gao2025, Tan2025}, while technical paradigms like RAG and 3D integration improve response accuracy and accessibility \cite{Zhao2024, Bodonhelyi2025}. However, general-purpose models still struggle with unnatural interactions and content reliability \cite{Zhang2025}.

While digital human in education have shown progress in domain-specific settings, including medical reasoning \cite{wang2025}, programming support \cite{Yang2024}, and neuroadaptive engagement \cite{Baradari2025}, systematic deployment guidelines for large-scale environments like MOOCs remain scarce.

\subsection{The Need for Digital Humans in Online Education}

AI-driven digital teachers address traditional online education flaws (such as poor support for massive learners and generic feedback) by enabling scalable, personalized interaction. Current systems like LittleMu \cite{tu2023} and PCA \cite{Hayashi2015} demonstrate massive reach (80k+ users) and improved concept explanation, yet often lack natural, human-like dialogue. While LLM-based solutions like VATE \cite{Xu2024AIDrivenVT} offer high-accuracy (78.3\%) and generative AI (GAI) systems  (e.g., ChatGPT) can aid writing feedback \cite{Han2024TeachersParentsStudents}, they face challenges regarding interactional rigidity, algorithmic bias, and authorship. Similarly, immersive VR agents \cite{Zhu2025, Petersen2021} and specialized tools like DesignQuizzer \cite{Peng2024} enhance trust and domain-specific knowledge but remain constrained by inconsistent experiences and limited adaptive guidance. Consequently, while digital teachers are vital for educational equity, significant gaps persist in cross-scenario adaptability and ethical management.

\subsubsection{The Importance of AI and LLMs in Online Learning.}

AI and LLMs redefine online education through automated lesson planning\cite{Fan2024}, real-time support \cite{Bodonhelyi2025}, adaptive assessment \cite{Chen2024}, and agent development \cite{Jin2025, Rogers2025}. High-precision systems like SAM (97.6\% accuracy rate in the chatbot’s responses) \cite{Bodonhelyi2025} and ChatTutor \cite{Chen2024} demonstrate advanced tutoring capabilities. Effective AI integration requires balancing automation with human oversight through targeted design. For instance, teachers utilize tools like TeachTune \cite{Jin2025} and LessonPlanner \cite{Fan2024} for agent review and adaptive planning, while learners engage with constrained systems like TeachYou \cite{Jin2024} for learning-by-teaching tasks. However, providing raw LLMs access without more tailored and scaffolded chatbot in low adoption rate, as shown in university courses where LLM-focused classes saw high usage \cite{Hellas2024}. Recent research explores AI and LLMs applications across diverse educational contexts, for example generative AI in writing education \cite{Han2024TeachersParentsStudents}, teacher training for K-12 AI curricula \cite{Jetzinger2024} and collaborative learning bots \cite{Cai2024}. Key implementation factors include real-time intervention \cite{Bodonhelyi2025}, content factors (knowledge restraint \cite{Jin2024}, adaptive generation \cite{Fan2024}, context awareness \cite{Bodonhelyi2025}), and intervention timing \cite{Cai2024}. Successful integration requires sophisticated design based on educational theory and practical constraints rather than simply deploying advanced language models.

\subsubsection{The Need for a New Framework for Online Learning with Digital Teachers.}

Existing online education research provides diverse frameworks for MOOC dimensions \cite{Amado2022MOOCsDesign}, scalable large-scale assessment \cite{Kasch2017}, and comprehensive pedagogy with multiple design principles \cite{Archambault03072022}, alongside specialized models for teacher competencies \cite{Saaiq02102024} and digital pedagogy integration \cite{Tan02072024}. However, these existing frameworks remain partial and fail to account for digital teacher scenarios involving autonomous, personalized instruction and adaptive tutoring at scale. 
AI-driven digital teachers offer several advantages over traditional tools: (1) real-time adaptation \cite{Han2024TeachersParentsStudents}; (2) continuous, large-scale personalized interaction through natural language processing \cite{Cai2024}; (3) the simultaneous integration of multiple functions \cite{Jetzinger2024}; and (4) extensive knowledge-based scaffolding \cite{Bodonhelyi2025}.
Implementations like Sara \cite{Winkler2020} and goal-oriented systems further demonstrate the significant efficacy of these intelligent pedagogical models.

\section{Methods} \label{sec:3}
This section details our research process, which began by establishing a conceptual design space and identifying key factors for digital teacher integration. To validate and refine these dimensions, we conducted two empirical studies gathering learners' and experts' insights on the framework in online learning. 

\subsection{Design Space and Key Factors} 

To address online learning’s multifaceted challenges, we synthesized 87 papers from education, HCI, and design to define the framework's design space and key factors. This structured approach establishes the core dimensions and bounds our research scope.

\subsubsection{Design Space.}
Using design space analysis \cite{MacLean01091991}, the research question was structured along three dimensions: when to use, what to teach, and how to design\cite{Elliott2017LivingSR,Xu2023}.

\paragraph{When to Use.} 

The "when" dimension includes service availability and delivery timing \cite{Petter2008ISsuccess, Wang2009InstructorAdoption, Yeh2022ChatbotGuidance}. \textbf{Availability} relies on high system quality and platform convenience to sustain learning outcomes \cite{Petter2008ISsuccess, Wang2009InstructorAdoption, AlAdwan2020MOOC-TAM}. \textbf{Delivery} utilizes manual (learner-initiated) or automatic (system-initiated) triggers \cite{Yeh2022ChatbotGuidance}. While manual triggers empower learner agency \cite{Roy2019Automation}, automatic just-in-time interventions (based on detected intent and context) significantly boost effectiveness \cite{NahumShani2018JITAI, Sun2021TeacherLMS}. To mitigate cognitive load, delivery must be adaptive rather than continuous \cite{Sadaf2019QMimpact, brainsci15020203}.

\paragraph{What to Teach.} The "what" dimension involves instructional content types and their delivery detail. \textbf{Content.}  Grounded in Bloom's Taxonomy and Gagne's Nine Events \cite{bloom1956taxonomy, anderson2001taxonomy, gagne1985conditions}, we define six content categories for digital teacher-enabled learning: (1) Orientation for course overviews and objectives; (2) Conceptual for factual and theoretical knowledge; (3) Procedural for skill acquisition and interactive simulations; (4) Affective for social and emotional learning; (5) Inquiry-based for critical thinking and exploratory projects; and (6) Evaluative for automated assessment and feedback mechanisms. 

\paragraph{How to Design.}
The "how" dimension focuses on modality and paradigm for digital teacher-enabled content representation. \textbf{Modality}. Digital teachers deliver multi-modal content (primarily text, visual, and audio) to enhance learning outcomes \cite{Guo2024DualView, Cuevas2017VisualAuditory, deOliveiraNeto2015AudioText}. While haptic or olfactory modalities remain constrained by technical bandwidth, the integration of text and audio provides a comprehensive foundation for educational applications. \textbf{Paradigm.} The visual paradigm encompasses various formats (textual, graphical, and video \cite{Xu2020CourseVideo, Simonyan2013DeepIC, Zeiler2014ECCV}) and three key patterns: interaction (e.g., click-based, drag-and-drop), presentation (e.g., dynamic vs. static visualizations), and spatial arrangement (e.g., layout and composition) \cite{Hinckley2004IODevices, Ragan2011, Min2025}. Conversely, the audio paradigm is more constrained, focusing on acoustic features such as volume, speed, and accent \cite{Frauenberger2007AudioUISurvey}.

\subsubsection{Key Factors.}

Another important aspect of the framework is the factors that determine the answers to the design space. We summarize these factors from two perspectives: digital teacher-specific and platform-agnostic.

\paragraph{
Digital Teacher-Specific Factors.}
We identify four key factors essential to the framework: digital teacher technology, learner state, course design, and contextual information.

\textbf{Digital Teacher Technology.}
Digital teacher technology provides personalized, on-demand support, overcoming the feedback delays inherent in traditional human instruction \cite{Zhao2024, Lin2023AIinITS, Graesser2001ITSdialogue}. This is achieved through LLM-powered chatbots for real-time debugging, neuroadaptive tutors that adjust content via EEG \cite{Baradari2025}, and 3D avatars utilizing retrieval-augmented generation (RAG) and emotional speech for humanized interaction \cite{Zhang2025, Zhao2024}. In VR environments, factors such as role-switching, appearance, and perceived social presence are critical to fostering learner trust and knowledge gain \cite{Petersen2021, Zhu2025}.

\textbf{Learner State.}
Digital teachers gain instant insights into learner states by analyzing multimodal analytics, physiological sensors, and system logs \cite{Acosta2024MLA, GIANNAKOS2019108, ALNASYAN2024100231}. Engagement and cognitive states are detected through webcam-based facial tracking, eye-tracking, and EEG data, alongside behavioral patterns like forum activity and daily engagement \cite{Kaur2018, Xie2023, GIANNAKOS2019108, Acosta2024MLA}. These detections enable real-time pedagogical adaptation, such as providing tailored dialogue and support for detected confusion or frustration, to enhance performance and reduce dropouts \cite{D'mello2013, KIZILCEC201718, ALNASYAN2024100231}.

\textbf{Contextual Information.}
The platform with AI teachers offers enhanced contextual capabilities compared to traditional e-learning, leveraging AI for highly adaptive and personalized experiences \cite{Gligorea2023AdaptiveAI ,Kabudi2021AIEnabled}. By perceiving mobile environments \cite{Gumbheer2022PersonalisedAdaptive} and utilizing context-aware recommenders \cite{Verbert2012ContextSurvey}, the framework facilitates individualized knowledge tracing, quiz-triggered content delivery, and location-aware resource recommendations \cite{Pardos2010KnowledgeTracing, DUPLOOY2024e39630, ElBishouty2007PERKAM}. This contextual depth allows rule-based mechanisms to dynamically adjust learning paths and instructional strategies according to the detected environment, device, and user profile \cite{Hasanov2019Survey}.

\textbf{Course Design.}
Digital teacher integration shifts course design from generalized pedagogical scaffolds \cite{Li_2024_06,Subramanian2020} toward dynamic, data-driven personalization \cite{Sghir_2022_12,Xie2025}. This paradigm emphasizes human-AI collaboration \cite{Xie2025}, personalized creative paths, and scalable LLM support \cite{Salminen2024}, while necessitating high interaction quality and proficient prompt engineering \cite{Mumtaz_2023_10, Zamfirescu-Pereira2023}. Ultimately, this redefines educators as "playwrights" who, supported by digital self-efficacy, script interactive learning experiences through no-code platforms \cite{Hedderich2024}.

\paragraph{Platform-Agnostic Factors.}

These factors integrate dual-perspective goals (learner and system) with comprehensive learner profiles encompassing prior knowledge, cognitive abilities, and learning preferences.

\textbf{Learner Goal.}
Learners pursue four primary goals during digital teacher-led pedagogical delivery: (1) knowledge mastery, supported by automated feedback, explanations, and structured learning paths toward mastery \cite{Kochmar_2020, Rus_2014}; (2) self-regulated learning (SRL), where digital teachers facilitate independent learning skills like goal-setting and help-seeking to predict and attain success \cite{WONG2021106913, Aleven_2016}; (3) AI literacy and collaboration, encompassing the understanding of AI mechanics and learning to partner with AI on complex tasks  \cite{Kim_2022, Jia_2025}; and (4) effective interaction, which focuses on responsive, trustworthy experiences tailored to individual demands to enhance performance \cite{Kumar_2024, Zhang_Li_2025}.

\textbf{Learner Profile.}
Learner profiles are essential for personalized design, integrating cognitive proficiency and metacognitive strategies within SRL beyond basic performance data \cite{Kay_2013, Gao_2025, KIZILCEC201718}. Such modeling allows a digital teacher to optimize learning trajectories, provide real-time assistance, and adapt to affective cues for enhanced engagement \cite{Loh_2021, Bodonhelyi2025, Park_2019}. Profiles are further enriched by multimodal physiological data (e.g., eye-tracking, EEG) and interaction patterns to improve performance prediction \cite{GIANNAKOS2019108, Russell_2025}. Finally, making these multifaceted profiles scrutable empowers students to reflect on and control their own learning process \cite{Kay_2013,Barria_Pineda_2018}.

\textbf{System Goal.}
Digital teachers pursue three fundamental goals to support online learning: (1) need identification and assistance, where systems monitor cognitive states (e.g., detecting mind wandering or confusion) and respond to direct inquiries for context-aware, personalized aid \cite{Bosch_2021, DMELLO2014153}; (2) responsiveness, characterized by constant availability to react to implicit behaviors or explicit questions through dynamic knowledge modeling and retrieval practice \cite{Levenberg_2023, Baillifard2023ImplementingLP}; and (3) enhanced efficiency, motivation, and retention, achieved through personalized practice schedules and instant feedback that boost satisfaction \cite{Baillifard2023ImplementingLP, Mohamed2025}. Furthermore, the use of embodied virtual avatars can significantly foster positive emotions and learning motivation, particularly when the digital teacher's appearance resonates with the learner \cite{Pataranutaporn_2022}.

\subsection{Study 1: Learner Survey} \label{sec:study1}

To understand how the unique features of a digital teacher might alter established findings on online learning, we surveyed learners to identify their preferences for this new educational experience.

\subsubsection{Participants and Recruitment.}
A total of 132 participants (primarily from urban Chinese contexts) completed the survey, providing a focused yet potentially localized perspective on digital teacher preferences. Their ages ranged from 18 to 66 years ($M = 25.19$, $SD = 5.34$). The sample consisted of 74 females (56.1\%) and 58 males (43.9\%). Most respondents were students, while others worked as technicians or in various other professions. Given participants’ varying levels of AI-related digital literacy, they were divided into four groups\footnote{No respondents identified as "unfamiliar with AI" during screening.}: (1) those who had worked on AI products ($n = 11$, 8.3\%), (2) those who used AI products frequently ($n = 92$, 69.7\%), (3) those who used AI products occasionally ($n = 25$, 18.9\%), (4) those who had heard of AI but never used it ($n = 4$, 3.0\%).

\subsubsection{Procedure.}
The survey utilized a mixed-methods approach to assess digital teacher effectiveness and design preferences. Quantitative Likert-scale and multiple-choice items evaluated learner motivation and support scenarios, while qualitative open-ended questions captured nuanced challenges and expectations. Following Braun and Clarke’s thematic analysis \cite{BraunClarke2012TA} (Table 1), two researchers independently coded the data using a collaboratively developed schema, resolving discrepancies through consensus. This dual-phase approach provided structured attitudinal measures alongside qualitative insights to inform design implications. Finally, the study was approved by the university's Institutional Review Board (IRB) to ensure full ethical compliance for research involving human subjects.

\begin{table}[!t] 
\vspace{-6mm} 
\centering
\small 
\caption{Insights from thematic analysis of survey.}
\label{tab:thematic analysis}
\renewcommand{\arraystretch}{1.1} 
\setlength{\tabcolsep}{3pt} 
\begin{tabularx}{\linewidth}{>{\bfseries}l >{\RaggedRight\arraybackslash}X >{\RaggedRight\arraybackslash}X >{\RaggedRight\arraybackslash}X} 
\toprule
Problem & Theme & Reason & Improvement Direction \\ 
\midrule

\multirow{3}{*}{\makecell[l]{MOOC \\ Dissatisfaction}} 
& Lack of interaction & No real-time communication, questions unanswered & Interactive features and timely feedback \\ \cmidrule(lr){2-4}
& Insufficient support & Few exercises, lack of feedback & Practice tasks and adaptive feedback \\ \cmidrule(lr){2-4}
& Monotonous experience & Flat pace, learners easily distracted & Diversify engagement techniques \\ 
\midrule

\multirow{2}{*}{\makecell[l]{AI \\ Dissatisfaction}} 
& Unstable quality & Responses inaccurate or fabricated & Improve reliability/transparency \\ \cmidrule(lr){2-4}
& Incomplete function & Unclear task boundaries, lack of templates & Structured tools/templates \\ 
\midrule

\multirow{2}{*}{\makecell[l]{Digital \\ Teacher}} 
& Limited usage & Many respondents had no real exposure & Increase accessibility/trials \\ \cmidrule(lr){2-4}
& Weak interaction & Few courses, complex operation & Simplify interface/interactivity \\ 
\midrule

\multirow{3}{*}{\makecell[l]{Needs}} 
& Expansive content & Expect advanced/key-topic explanations & Provide critical content \\ \cmidrule(lr){2-4}
& Interactive & Request time-sensitive guidance & Enable real-time explanations \\ \cmidrule(lr){2-4}
& Domain relevant & Need tailored content for specific domains & Incorporate domain cases \\ 
\bottomrule
\end{tabularx}
\vspace{-6mm} 
\end{table}

\subsubsection{Results.}
The survey revealed that participants have clear preferences regarding the timing, content, and modality of instructional support (e.g., teaching, assistance and evaluation).

\begin{figure}[htbp]
  \centering
  \begin{minipage}{0.32\linewidth}
    \centering
    \includegraphics[width=\linewidth]{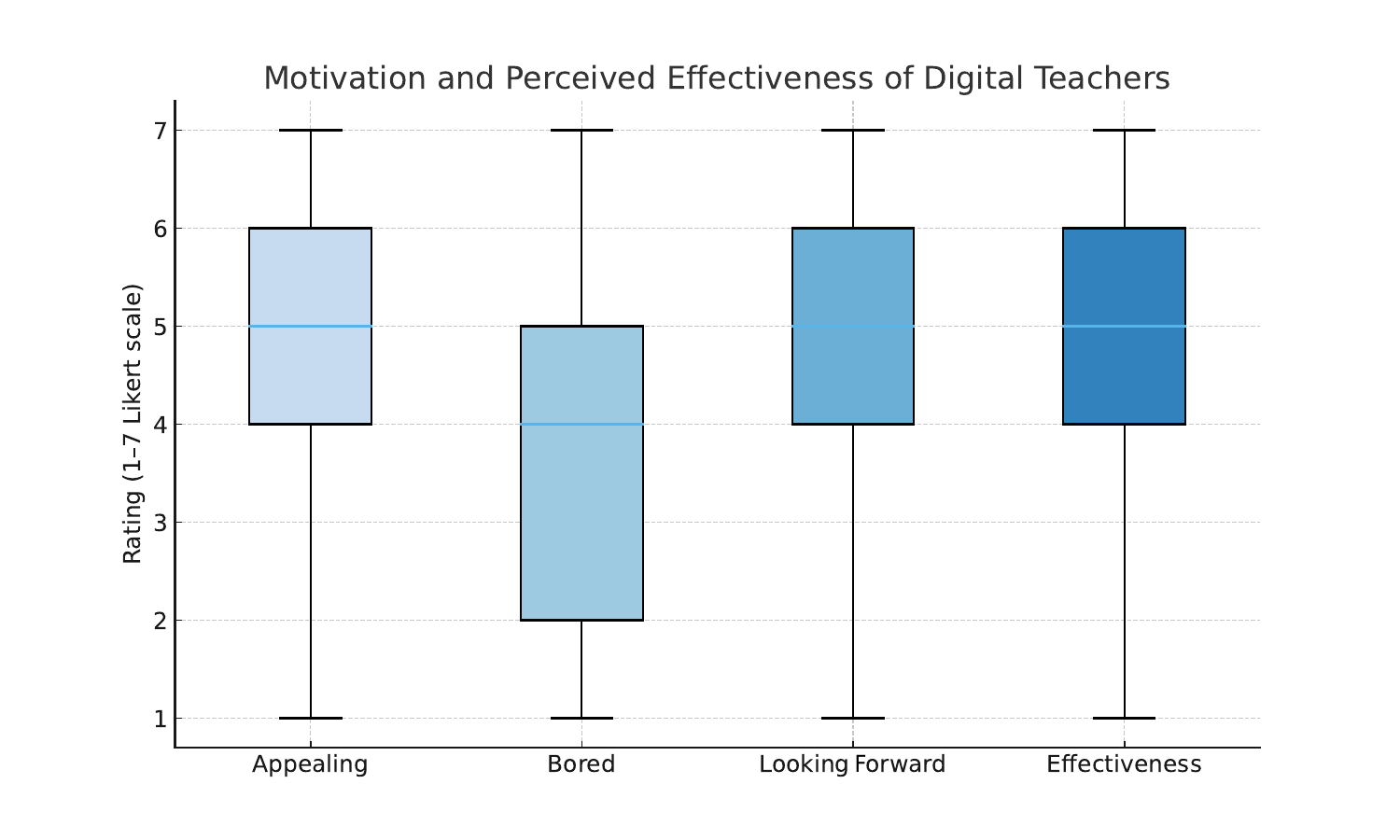}
    \caption{Motivation and perceived effectiveness of digital teachers.\centering}
    \label{fig:motivation-dt1}
  \end{minipage}
  \hfill
  \begin{minipage}{0.32\linewidth}
    \centering
    \includegraphics[width=\linewidth]{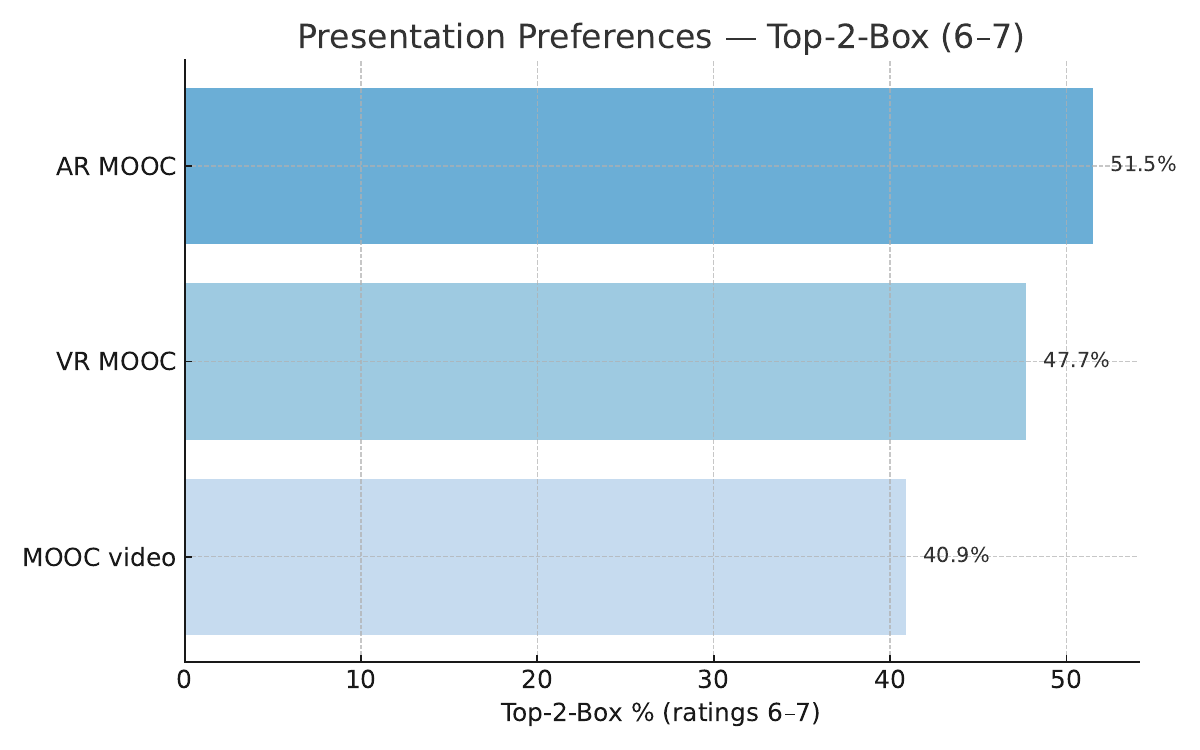}
    \caption{Preference for immersive modalities in digital teachers.\centering}
    \label{fig:motivation-dt2}
  \end{minipage}
  \hfill
  \begin{minipage}{0.32\linewidth}
    \centering
    \includegraphics[width=\linewidth]{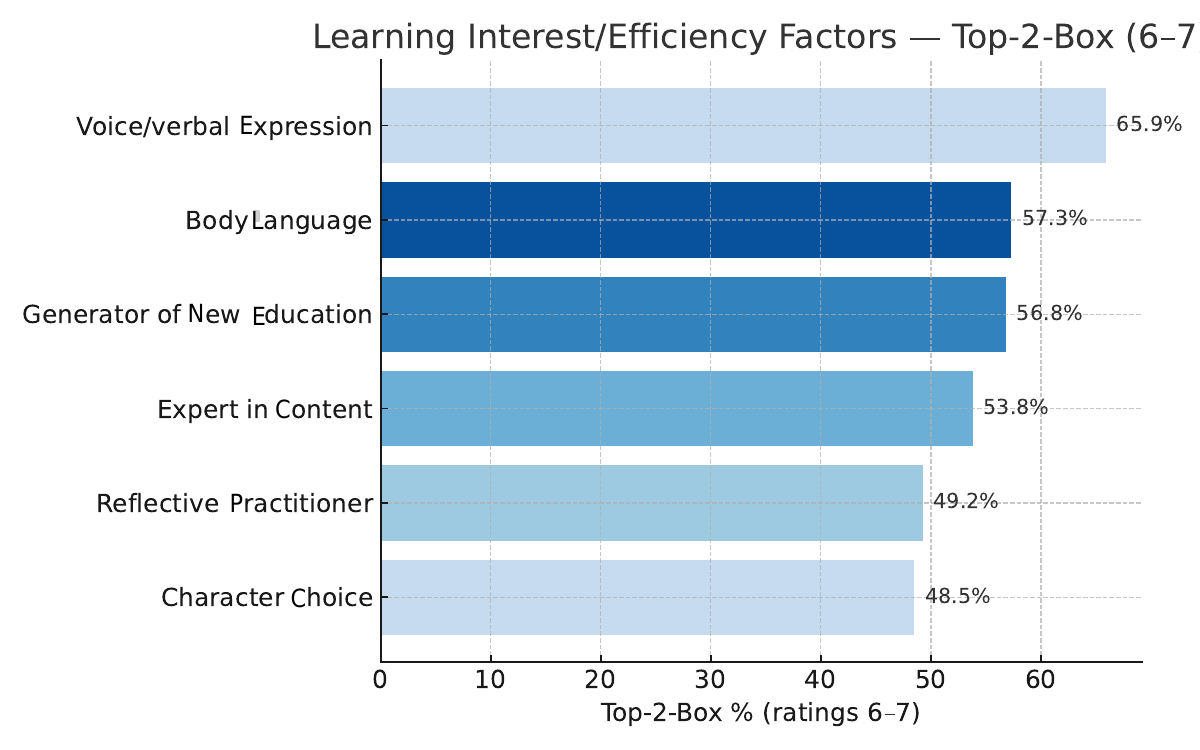}
    \caption{Preference for expressive modalities in digital teachers.\centering}
    \label{fig:motivation-dt3}
  \end{minipage}
  \vspace{-2em} 
\end{figure}

\textbf{Finding 1: Most learners expressed a desire for digital teachers to provide teaching, assistance, and evaluation within MOOCs} (related to when - availability). 
Survey results indicate a strong desire for digital teachers to provide comprehensive teaching, assistance, and evaluation within MOOCs. While most participants have MOOC experience ($n=94, 71.2\%$; satisfaction: $76.6\%$) and positive sentiments toward AI products ($89.1\%$), digital teacher usage remains low ($15.2\%$) with mixed satisfaction ($60.9\%$). Likert ratings (1-7 scale) reflect a positive outlook, with medians of 5 for appeal, expectation, and perceived effectiveness; however, boredom (Median $= 4$) remains a critical area for improvement \textbf{(see Fig. \ref{fig:motivation-dt1})}. Qualitative feedback identified two primary pain points: poor avatar/UI quality [C1, C7] and stiff, formulaic interaction [C2-6].

\textbf{Finding 2: The majority of learners wanted assistance from the digital teacher on an as-needed basis} (related to when - delivery). 
The vast majority strongly agreed that the digital teacher should "often" assist them ($77.8\%$ at score 6-7; 1-5: $22.2\%$), while "rarely help" was largely rejected ($62.8\%$ at scores 1-2). Support "when necessary" was viewed positively (peak 5-6: $44.7\%$) but less emphatically (score 7: $15.2\%$). Qualitative feedback on dissatisfaction with AI centered on three themes: (1) Accuracy, citing hallucinations and errors [B3,4,9,11]; (2) Functional limits, involving generic, non-customized responses [B2,3,6,7,11-13]; (3) Usability, noting inefficiency and technical immaturity [B5,8,14].

\textbf{Finding 3: Learners expressed a clear preference for specific types of instructional support} (related to what - content). Survey results reveal a strong preference for conceptual and inquiry-based teaching materials, with $70.5\%$ and $70.4\%$ agreement respectively. Evaluative materials also scored high at $69.7\%$. Procedural and Orientation materials were favored by $58.4\%$ and $66.6\%$, respectively. Affective content was less prioritized, with only $33.4\%$ agreement. This suggests that while learners recognize the importance of affective (social and emotional) learning, they prioritize cognitive and practical skills development more.

\textbf{Finding 4: Learners expressed a strong preference for expressive and immersive modalities} (related to how - modality). Regarding presentation modes, $51.5\%$ placed AR-based MOOCs among the top categories, followed by VR-based MOOCs ($47.7\%$), whereas only $40.9\%$ expressed similar enthusiasm for digital teachers embedded in traditional MOOC videos \textbf{(see Fig. \ref{fig:motivation-dt2})}. Concerning factors that enhance learning interest and efficiency, the most salient was voice and verbal expression ($65.9\%$), followed by body language ($57.3\%$) and expert in content ($53.8\%$). In contrast, roles such as reflective practitioner ($49.2\%$) and  appearance/character choice ($48.5\%$)were rated moderately. These findings highlight that learners prioritize expressive and immersive qualities of digital teachers over conventional delivery modes \textbf{(see Fig. \ref{fig:motivation-dt3})}. 

\textbf{Finding 5: Interactivity, personalization, and usability are crucial design priorities for future digital learning systems.} 
Qualitative insights (see Table \ref{tab:thematic analysis}) highlighted critical gaps in current systems. For MOOCs, learners emphasized a lack of real-time interaction and limited support, desiring diversified instructional approaches. AI product dissatisfaction stemmed from poor domain adaptation and unstable output quality, necessitating reliable, specialized models. Regarding digital teachers, limited usage experience and interface complexity drove demands for simplified usability and expanded content. Finally, requirements for detailed explanations centered on forward-looking, domain-relevant, and timely guidance. Overall, these findings underscore interactivity, personalization, and usability as essential pillars for future digital learning design.

\subsection{Study 2: Co-design Workshops}  
Based on literature and survey data, an initial design framework was developed and subsequently refined through three stakeholder workshops (learners and experts). We adopted a qualitative approach to explore digital teacher integration in online learning (e.g., MOOCs), specifically examining usage patterns, attitudes, and future expectations to optimize the design process and learner experience.

\begin{figure}[htbp]
  \centering

  \begin{minipage}[t]{0.32\linewidth}
    \centering
    \includegraphics[width=\linewidth,height=3.5cm]{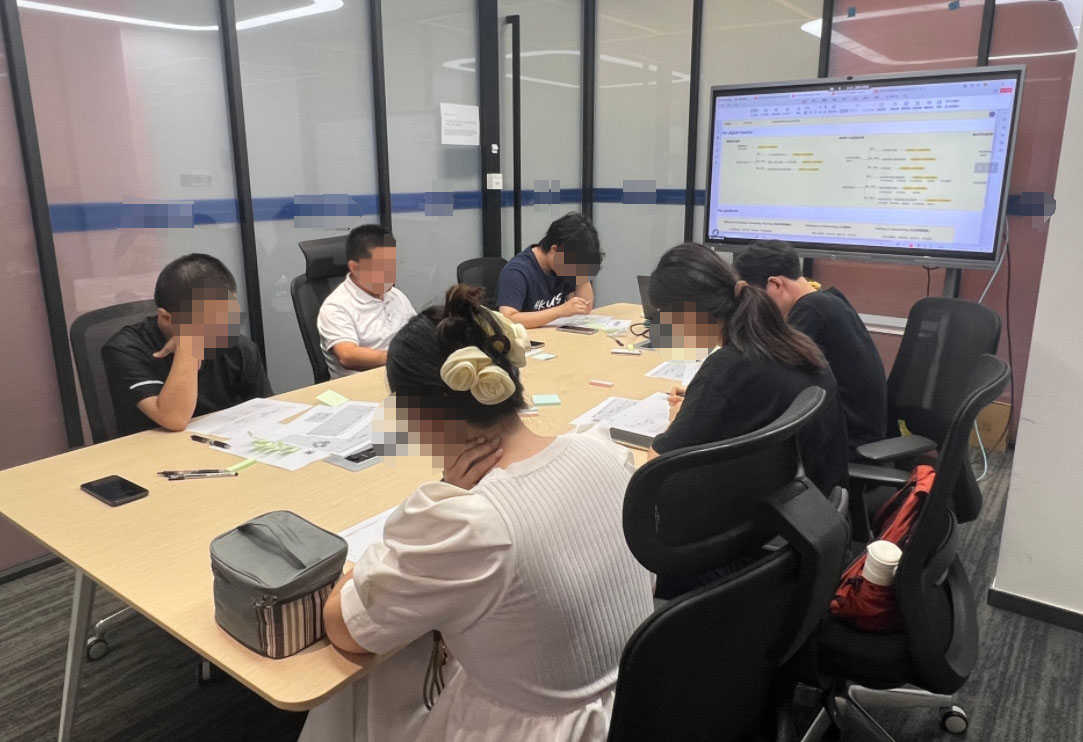}
    \par\small (a) Workshop Session in Progress.
  \end{minipage}\hfill
  \begin{minipage}[t]{0.32\linewidth}
    \centering
    \includegraphics[width=\linewidth,height=3.5cm]{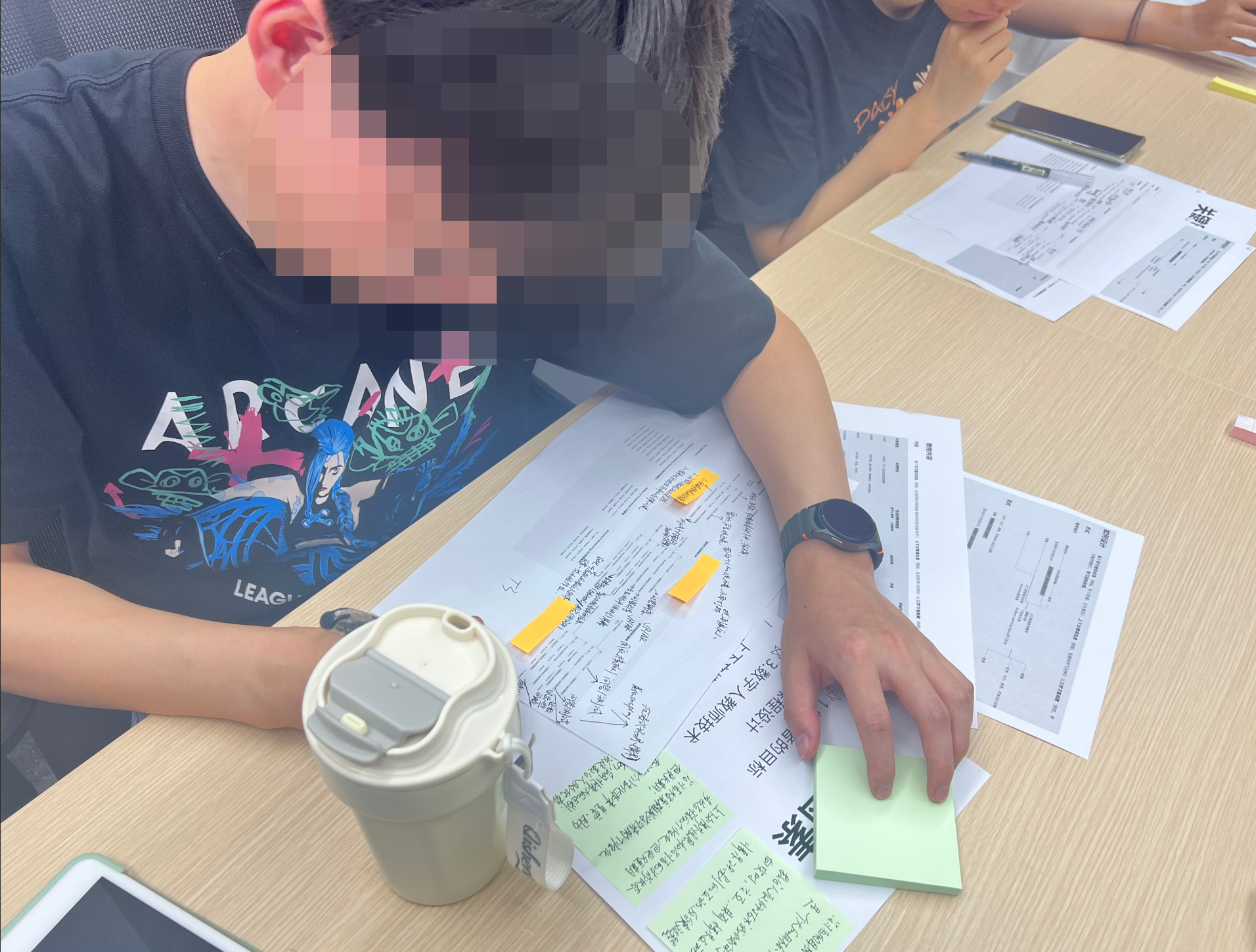}
    \par\small (b) Collaborative Ideation.
  \end{minipage}\hfill
  \begin{minipage}[t]{0.32\linewidth}
    \centering
    \includegraphics[width=\linewidth,height=3.5cm]{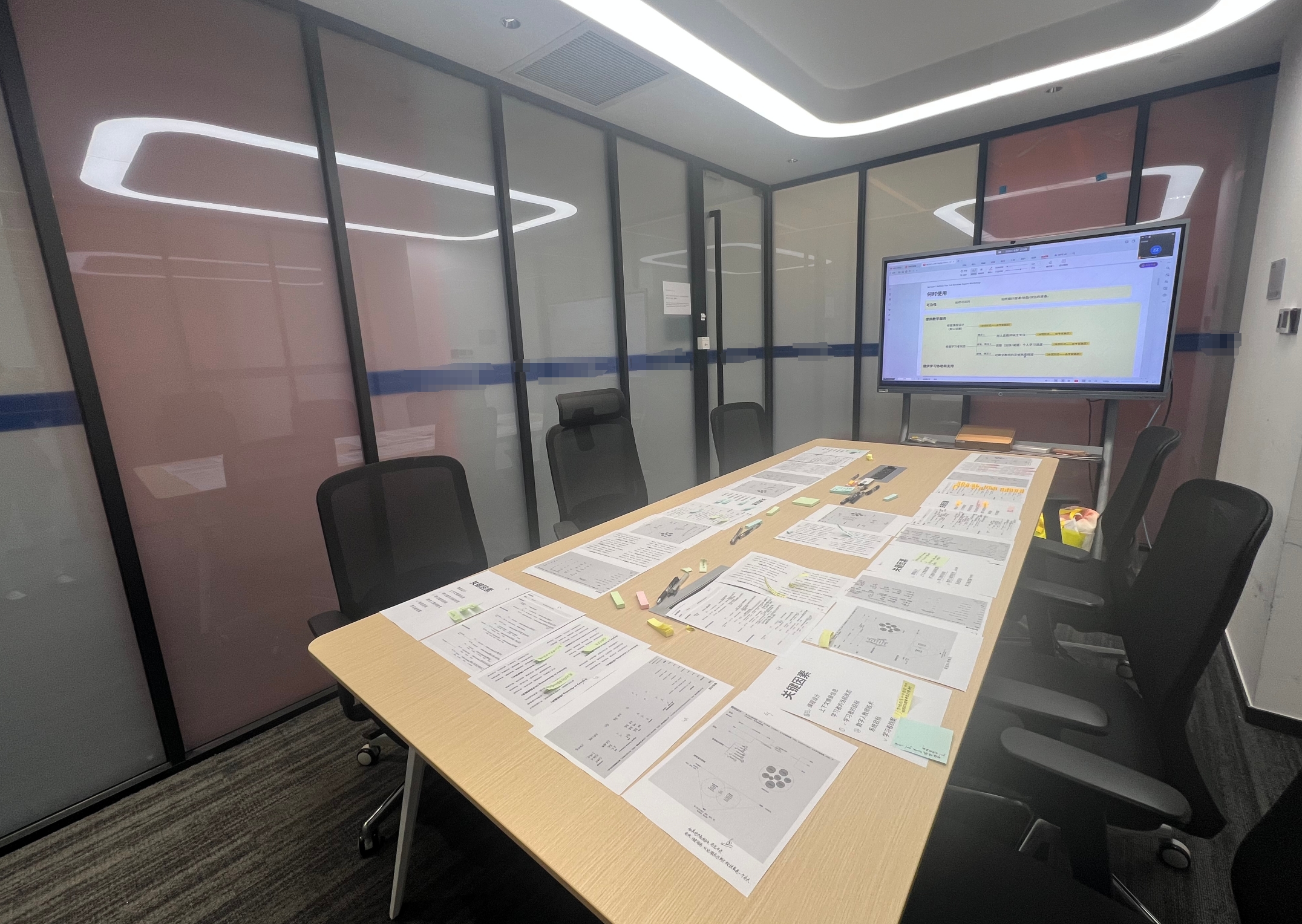}
    \par\small (c) Experimental Environment Setup.
  \end{minipage}

  \caption{Procedural overview of study 2.}
  \label{fig:workshop}
  \vspace{-2em}
\end{figure}

\subsubsection{Participants.}
We recruited eighteen participants ($11$female, $7$ male; $M=25.41$, $SD=3.5$) from a technology university. The participants represented four domains: technician ($n=3$), educator ($n=6$), user experience/interface designer (UX/UI designer) ($n=3$), and learner ($n=6$). 
All of them were familiar with AI and digital teacher concepts. For the study, they were randomly assigned to one of three balanced, interdisciplinary groups, each comprising one or two expert from every domain.

\subsubsection{Procedure.}


Pre-workshop materials provided context on digital teachers for expert participants. Sessions utilized a large screen and sticky notes \textbf{(see Fig. \ref{fig:workshop})}, where we presented a draft framework synthesized from literature and survey findings. To facilitate rich, in-place feedback, participants directly annotated printed Chinese versions of the framework documents.



\begin{figure}[htbp]  
\vspace{-1.5em}
  \centering
  \includegraphics[width=1\linewidth]{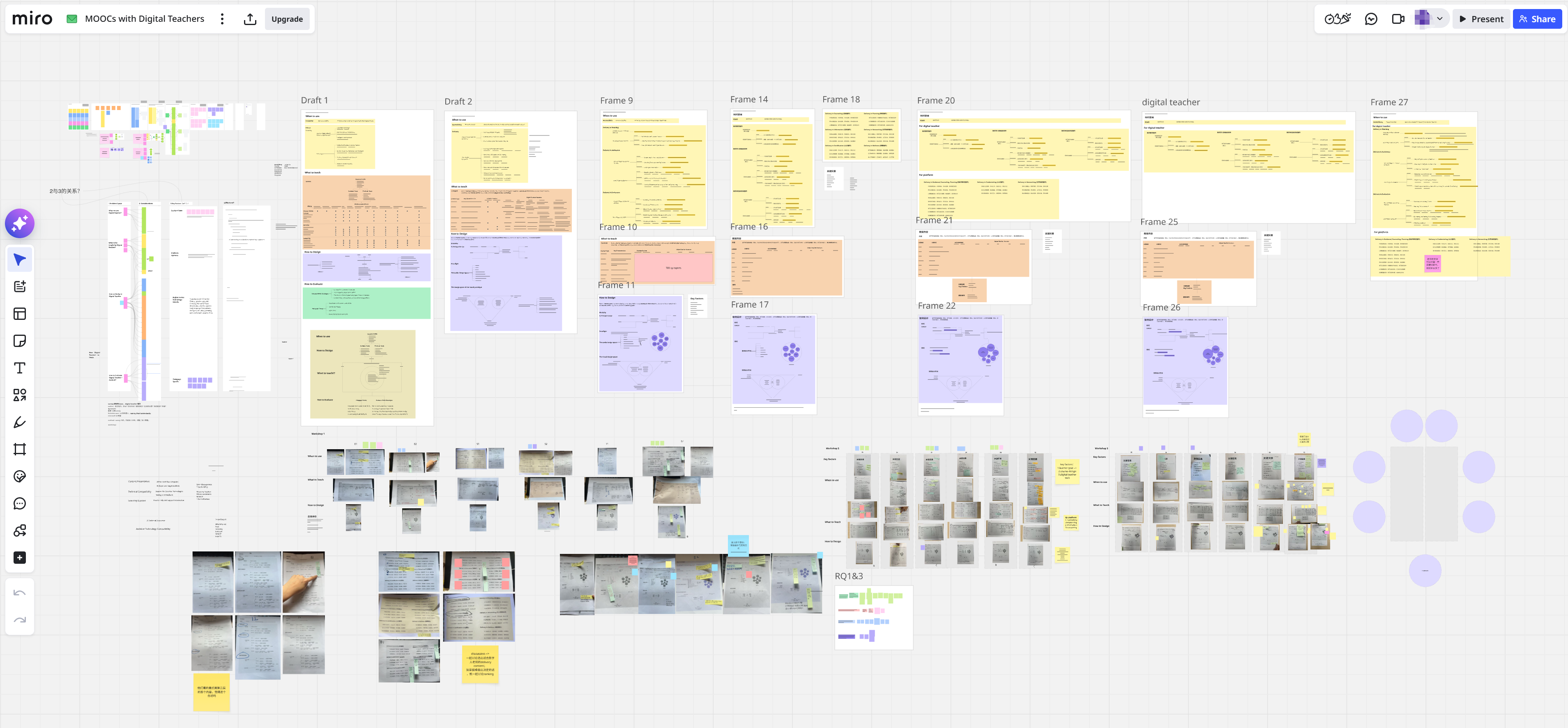}
  \caption{Affinity diagram analysis process for iterative framework development.}
  \label{fig:affinity}
  \vspace{-2em}
\end{figure}

Participants provided informed consent with guaranteed anonymity and received 100 CNY/hr compensation. The study followed an iterative cycle across three recorded workshops (60–150 min each). After each session, two researchers applied affinity diagramming \cite{Harboe2015} to analyze data and iteratively refine the framework for subsequent stages \textbf{(see Fig. \ref{fig:affinity})}.

\subsubsection{Results.}


Experts participants highly recommended the framework, describing it as "very useful" (E1, S1-2, D1, D3, T3). Feedback evolved from broad conceptual suggestions in early workshops to granular, specific insights as the framework reached convergence. Key refinements are categorized below:

\textbf{Recommendation 1: Add Missing Parts.}   
Participants identified missing parts, leading to defined presentation formats for the "When to Teach" dimension and an emphasis on the digital teacher's strengths in guidance, information, and networking (E4). Specifically, the "Modality" dimension was expanded to incorporate real-time learner state adaptivity (E2, T2) and avatar customization (S1, S2), while the "Paradigm" dimension was updated to include explicit interaction design principles (T1). These insights underscore the necessity for a practical, actionable guide that prioritizes a dynamic, learner-centered design approach.

\textbf{Recommendation 2: Reduce Unnecessary Content.} 
Participants also found some parts of the framework to be overly complex or redundant. For example, following the third workshop, the experts reached a consensus to streamline the design space of audio, condensing it into three core components: accessibility, customized selection, and adaptive control.

\textbf{Recommendation 3: Focus on What Matters Most.} 
Group 3 identified learner goal, learner profile (n=4), and digital teacher technology (n=3) as foundational elements. To address inherently subjective learner preferences, the digital teacher serves as an ideal medium that, mitigating current technical limitations, provides personalized instruction tailored to individual needs.

\textbf{Recommendation 4: Improve Integration and Visual Structure.} 
Participants unanimously supported D3's proposal for a closed feedback loop. This mechanism leverages learner feedback (e.g., via questionnaires) to continuously optimize system input, facilitating a iterative refinement process for multimodal content, including text, audio, and images - to enhance subsequent interactions.

\section{Framework}  


Following the initial framework structure and empirical results, we synthesize literature with our study’s findings and recommendations. This integration details the complete digital teacher framework, mapping how key factors inform design choices, addressing the when, what, and how, to establish an actionable framework.

\subsection{When to Teach}

\subsubsection{Availability.} Survey results highlight significant demand for digital teachers in online learning environments (e.g., MOOCs) (Finding 1). Systems must provide constant pedagogical support (teaching, assistance, and evaluation) to ensure assistance is available at the learner's "moment of need" \cite{Zanzotto_2019}.


\subsubsection{Delivery.}


Aligned with previous work \cite{brainsci15020203,Yeh2022ChatbotGuidance}, the default approach should be to wait for learners to manually request support. Building on this, our framework divides "When to Teach" into platform and digital teacher perspectives. 

\paragraph{The Digital Teacher Dimension} For the Digital Teacher dimension, experts made recommendations regarding the When to Teach component, specifying the presentation formats for the three distinct supports a digital teacher can deliver: teaching, assistance, and evaluation.

\textbf{Delivery in Teaching.} 
Teaching delivery is determined by: (1) Course Design (default), utilizing integrated multimedia (video, text, audio); and (2) Learner State, where the system provides haptic/auditory alerts for loss of focus, manages pacing via quizzes for progress adjustments, and employs dynamic profiling for experienced users.

\textbf{Delivery in Assistance.} 
Triggers are either Manual (default) or Automatic. Manual assistance supports (a) learning guidance and (b) curiosity-driven exploration through Q\&A sessions (e.g., immersive VR/AR), or (c) looking for interaction with digital teachers to test their knowledge. Automatic assistance addresses (a) learner confusion through context-aware multimedia, (b) attention monitoring via test results, physiological data and interactive avatars, and (c) collaborative learning (e.g., through Socratic dialogue) with virtual partners based on Learner goals and profiles.

\textbf{Delivery in Evaluation.}
Evaluation follows Manual or Automatic mechanisms. Manual evaluation includes (a) process assessment via templates and (b) self-assessment through Q\&A or AR/VR role-playing, such as T3’s suggested poetry recitation scenario, informed by learner state. Automatic evaluation focuses on (a) learning outcomes through tailored exams, (b) progress monitoring via visual indicators and conversational Q\&A, and (c) feedback delivery through systematic summaries, data visualizations, and sound effects tailored to contextual factors.

\paragraph{The Platform Dimension}

Experts expanded the platform dimension to include Guidance, Credentialing, and Networking, citing these as unique digital teacher strengths. Delivery modes include: (1) Guidance, covering admissions and career development counseling; (2) Credentialing, managing competency records and professional certification guidance; and (3) Networking, facilitating the establishment and expansion of academic and professional networks.


\subsection{What to Teach}


\subsubsection{Content.} 


Instructional outputs are driven by six key factors: learner state, course design, contextual information, system goal, learner profile, and digital teacher technology. Our framework maps these factors to six content types derived from the literature. Experts reached a consensus that digital teachers excel in delivering conceptual, procedural, and evaluative content, although inquiry-based tasks remain better suited for humans. Regarding affective content, opinions were split: while some preferred human empathy, T3 argued that digital teachers offer a "privacy advantage," making this type suitable for both.


\subsubsection{Details.} Based on our literature review, survey results, and expert advice, a clear conclusion emerges: learners find detailed explanations to be both useful and necessary. Elaborating on this point, the experts recommended that the instructional details provided by a digital teacher must be abundant, high-quality, and personalized.



\subsection{How to Design}

\subsubsection{Modality.}


Digital teacher modalities are driven by learner state, profile, goals, and contextual information. The design integrates foundational assets (text, audio, images) via interaction design principles, employing static personalization (profile-based) and real-time adaptation. This AI-driven process generates rich, multimodal outputs (e.g., animations, 3D models). Crucially, a closed-loop mechanism facilitates self-iterative improvement through continuous learner feedback to optimize future educational outcomes.


\subsubsection{Paradigm.}
The paradigm for a digital teacher includes two primary areas: the visual and audio design spaces. The visual space comprises textual (typography: color, size), video, and graphic (scenario-prioritized) formats. Interaction principles, such as clicks, scenario-based engagement, and abstract concept demonstrations, enhance these formats. Experts prioritized textual content and video over graphics, noting that integrating these with course-specific designs creates a cohesive digital teacher experience. Conversely, audio is deemed secondary, focusing on accessibility, customized selection, and adaptive control.





Our framework for human-AI pedagogical integration synthesizes empirical findings with multidisciplinary literature from online education and HCI, offering actionable guidelines for AI-driven instructional design.


    
    
    



\section{Discussion}


 Our human-AI pedagogical framework defines the digital teacher design space by mapping critical factors to instructional dimensions. As both a theoretical reference and practical toolkit, it guides the development of effective AI-driven educational solutions. We discuss our findings regarding stakeholder perceptions, the design space, and key factors. Finally, we outline the limitations of this work and propose future research directions inspired by our studies.
 
\subsection{Discussion 1: Comparison and Perception} 

Our study addresses RQ1 by highlighting a multifaceted comparison and perception of digital teachers in online learning experiences (i.e., MOOCs), synthesized from expert and learner feedback (Fig. \ref{fig:affinity}).

\subsubsection{Advantages of Digital Teacher Integration.} 
Based on the expert and learner feedback (Fig. \ref{fig:affinity}), we identified four primary advantages of integrating digital teachers into the online learning process: \textbf{(1) Information Processing Superiority}. Unlike humans, AI can rapidly collect, summarize, and organize internet-scale data, providing real-time information synthesis (E1, S3, D1). \textbf{(2) Operational Efficiency}. Digital teachers transcend human physiological limits through multitasking, superior memory, and consistent monitoring that exceeds human visual capabilities (D1). \textbf{(3) Personalized Support}. They enhance the learning experience through customized learner profiles, intuitive interaction design, and planning services that improve student time management (E3, D2). \textbf{(4) Psychological Safety and Niche Support}. Digital teachers provide a private, judgment-free environment for sensitive topics (T3) and fill the void where suitable human experts are unavailable for specific niche subjects (S3).


\subsubsection{Critical Challenges and Disadvantages.}
Despite their advantages, the integration of digital teachers presents several critical challenges categorized into three primary dimensions: \textbf{(1) Information Reliability and Specialization}. A major concern involves the authenticity and accuracy of AI-generated content. Participants noted risks of hallucination and lack of transparency regarding sources (B3, B4, B11, E1). Furthermore, professional expertise is difficult to guarantee in high-stakes or specialized fields, such as medicine (D2). \textbf{(2) Interactional Rigidity}. Unlike human educators, digital teachers are frequently perceived as formulaic, patterned, and rigid in their responses (B12, C2--C6). This lack of interactional flexibility often results in a "stiff" persona that fails to adapt to the nuances of complex student inquiries. \textbf{(3) Production Quality and Cognitive Distraction}. The visual and technical execution of the digital human significantly impacts learning. Poor production quality or uncanny digital human creation can become a distraction for students rather than an aid (C7). Additionally, usability issues such as poor UI navigation (C1) and efficiency problems during prolonged exploration (B8) can hinder the overall learning experience. \textbf{(4) Functional and Contextual Gaps.} Current AI implementations struggle with contextual memory, often "forgetting" previous dialogue or failing to handle nuanced tasks (B7, B13). Limitations in multi-modal capabilities (such as file uploads or image recognition) further restrict their utility in complex pedagogical scenarios (B14).


\subsubsection{Perceptions of Learners and Teachers.} 
The integration of digital teachers into MOOCs reveals a complex interplay between high functional expectations and deep-seated trust barriers. While the framework received strong endorsement from experts as "very useful" (E1, S1--S2, D1, D3, T3), the transition from traditional to AI-led pedagogy is marked by several critical considerations. \textbf{(1) The Trust-Utility Paradox}. A "cognitive lag" exists in AI education. While stakeholders recognize AI’s ubiquity and 24/7 efficiency (E1, E3, D1), a fundamental trust gap remains (E2). Skepticism regarding fabricated content leads to an attitude of cautious optimism: AI is viewed as a supplement to human limits, not a replacement (S2). \textbf{(2) From Technical Novelty to Pedagogical Substance}. Stakeholders now prioritize teaching effectiveness over technical novelty (S4). They demand real-time interaction and timely feedback to alleviate boredom (A6, A10, A15, A16). Learners also want the power to customize the digital teacher’s appearance for different educational contexts (S3). Learners express significant concern regarding the value and recognition of course certifications (E3). \textbf{(3) Trust and Privacy}. A significant "trust gap" remains, as stakeholders note that the mindset transition toward accepting digital humans as credible educators requires time (E2). Paradoxically, this lack of human presence offers a "privacy advantage": digital teachers provide a safe, judgment-free environment that enhances psychological safety, particularly for sensitive or personal learning topics (T3).


\subsection{Discussion 2: Design Space, Key Factors and Framework}

\subsubsection{Framework Development.}

Developing an effective framework for online learning is a multifaceted challenge. To address RQ2 and bound our research scope, we established a design space and key factors through a review of over 87 papers across education, HCI, and design research. This structure was further refined through empirical insights from our learner survey and expert workshops. Design decisions within this space are governed by two categories of key factors: \textbf{(1) Digital Teacher-Specific Factors}, which distinguish AI agents from human instructors, including teaching technology, real-time learner state, and contextual information; and  \textbf{(2) Platform-Agnostic Factors}, which encompass broader educational objectives such as system goals, learner goals, and learner profiles. Together, these dimensions and factors constitute the foundational structure of our integrated framework.

\subsubsection{Framework Application.}

Researchers and designers can leverage this framework to develop educational platforms (e.g., MOOCs) through an iterative, three-stage process. First, they conceptualize initial designs based on pedagogical intuition. Second, they apply the framework to identify seven critical factors-digital teaching technology, learner state, course design, and contextual information (teacher-specific), alongside system goals, learner goals, and learner profiles (platform-agnostic). Third, using the When, What, and How design space, they systematically inspect and modify these designs to address any gaps or functional misalignments.

\subsection{Discussion 3: Limitations and Future Directions}
There are a few limitations to this research. Our framework currently lacks empirical validation in more specific scenarios and operational guidance for abstract factors like learner state. Furthermore, we also should engage with the ethical considerations for using digital teachers. Strict privacy rules are needed so real-time sensing supports learning without compromising student data confidentiality. Ethically, digital teachers should supplement human educators in instructional shortcomings (S2) rather than replacing empathetic roles (T3). Future research will transition toward providing measurable factor definitions—mapping learner state to data points like quiz scores—and validating guidelines via end-user studies. As sensing technology evolves, we envision automating factor identification to foster real-time, self-optimizing, and human-centered pedagogical environments.

\section{Conclusion}
In this paper, we propose \textbf{the first framework} for the design and integration of digital teachers in online education. Drawing from a multidisciplinary literature review, we structure the design space around three core questions: when to use, what to teach, and how to design. We then developed the final framework 
by synthesizing findings from a survey with 132 learners (Study 1) and a series of iterative workshops with 18 experts (Study 2). As a contribution, the framework provides a structured approach for designers and a valuable reference for researchers, aiming to guide the creation of more effective digital teachers and to help explore new design opportunities.

\begin{credits}
\subsubsection{\ackname} 
This work was supported in part by the National Key Research and Development Program of China (Grant No. 2024YFC3307602), the Guangdong Provincial Talent Program (Grant No.2023JC10X009), and the Red Bird MPhil Program at the Hong Kong University of Science and Technology (Guangzhou).
\vspace{-10pt}

\subsubsection{\discintname}
The authors have no competing interests to declare that are relevant to the content of this article. 
\end{credits}
\vspace{-10pt}
%


%
%

%





%
%

%

\end{document}